 \documentclass[final,5p,times,twocolumn]{elsarticle}

\usepackage{amssymb}
\usepackage{lipsum}

\usepackage{bm}
\usepackage{amsfonts}
\usepackage{latexsym}
\usepackage[utf8]{inputenc}
\usepackage{graphicx}
\usepackage{amsmath}
\usepackage{textcomp}
\usepackage{subfigure}
\usepackage{commath}
\usepackage{caption}
\usepackage{multirow}
\usepackage{ulem}
\usepackage{tabularx}
\usepackage{booktabs}  
\usepackage{xcolor}
\usepackage{hyperref}

\usepackage{float}
\usepackage{dcolumn}
\usepackage{ragged2e}
\usepackage{hyperref}
\usepackage{tensor}
\hypersetup{colorlinks,citecolor=blue}

\journal{High Energy Astrophysics}

\begin{document}

\begin{frontmatter}



\title{Comparative analysis of machine learning techniques for feature selection and classification of Fast Radio Bursts}


\author[ufes1]{Ailton J. B. Júnior}
\ead{ailton.brandao@edu.ufes.br}

\author[uct]{Jéferson A. S. Fortunato\corref{cor1}}
\ead{jeferson.fortunato@uct.ac.za}

\author[ufes1]{Leonardo J. Silvestre}
\ead{leonardo.silvestre@ufes.br}

\author[ufes2]{Thonimar V. Alencar\corref{cor2}}
\ead{thonimar.souza@ufes.br}

\author[ufes2,ufes3]{Wiliam S. Hipólito-Ricaldi\corref{cor3}}
\ead{wiliam.ricaldi@ufes.br}

\affiliation[ufes1]{%
  organization={Departamento de Computação e Eletrônica, CEUNES, Universidade Federal do Espírito Santo}, addressline={Rodovia BR 101 Norte, km 60},
  city={São Mateus},
  postcode={29940-540},
  state={ES},
  country={Brazil}
}

\affiliation[uct]{%
  organization={High Energy Physics, Cosmology \& Astrophysics Theory (HEPCAT) Group, Department of Mathematics and Applied Mathematics, University of Cape Town},
  city={Cape Town},
  postcode={7700},
  country={South Africa}
}

\affiliation[ufes2]{%
  organization={Departamento de Ciências Naturais, CEUNES, Universidade Federal do Espírito Santo},
  addressline={Rodovia BR 101 Norte, km 60},
  city={São Mateus},
  postcode={29940-540},
  state={ES},
  country={Brazil}
}

\affiliation[ufes3]{%
  organization={Núcleo Cosmo-UFES, CCE, Universidade Federal do Espírito Santo},
  addressline={Av. Fernando Ferrari, 540},
  city={Vitória},
  postcode={29075-910},
  state={ES},
  country={Brazil}
}

\begin{abstract}
Fast Radio Bursts (FRBs) are millisecond-duration radio transients of extragalactic origin, exhibiting a wide range of physical and observational properties. Distinguishing between repeating and non-repeating FRBs remains a key challenge in understanding their nature. In this work, we apply unsupervised machine learning techniques to classify FRBs based on both primary observables from the CHIME catalog and physically motivated derived features. We evaluate three hybrid pipelines combining dimensionality reduction with clustering: Principal Component Analysis (PCA) + k-means, t-distributed Stochastic Neighbor Embedding (t-SNE) + Hierarchical Density-Based Spatial Clustering of Applications with Noise (HDBSCAN), and t-SNE + Spectral Clustering. To identify optimal hyperparameters, we implement a comprehensive grid search using a custom scoring function that prioritizes recall while penalizing excessive cluster fragmentation and noise. Feature relevance is assessed using principal component loadings, mutual information with the known repeater label, and permutation-based F\textsubscript{2} score sensitivity. Our results demonstrate that the derived features, including redshift, luminosity, and spectral properties, such as the spectral index and the spectral running, significantly enhance the classification performance. Finally, we identify a set of FRBs currently labeled as non-repeaters that consistently cluster with known repeaters across all methods, highlighting promising candidates for future follow-up observations and reinforcing the utility of unsupervised approaches in FRB population studies.
\end{abstract}



\begin{keyword}
Fast Radio Bursts \sep Unsupervised Learning \sep Astrophysical Feature Extraction



\end{keyword}

\end{frontmatter}




\section{Introduction}
Fast Radio Bursts (FRBs) are among the most intriguing phenomena in modern astrophysics. These brief, highly energetic radio pulses, typically lasting only a few milliseconds, originate from extragalactic distances and exhibit high dispersion measures (DM). The DM provides
information about the physical  properties of the intervening
medium and often exceeds the expected contribution from the Milky Way, offering strong evidence for their extragalactic origin. The first FRB was discovered in 2007 by Duncan Lorimer and collaborators \cite{lorimer2007bright}. Despite significant observational efforts, the physical mechanisms underlying FRBs remain uncertain. A particularly compelling observational dichotomy has emerged between repeating and non-repeating FRBs {~\cite{zhang2023thephysics}}, suggesting different progenitor scenarios or environmental conditions.

The short duration of the pulses, constrained by the light-crossing time of the emission region, implies that the source is extremely compact. The most plausible sources include highly magnetized neutron stars, black holes, and white dwarfs, each embedded in distinct astrophysical environments \cite{petroff2019fast, beniamini2023hybrid}. Other possible progenitors involve binary interactions, collisions, tidal disruption events, and potentially even exotic physics \cite{popov2020origin, xing2024short, pham2024fast, vachaspati2008cosmic}. So far, 862 FRB sources have been published by multiple collaborations, 69 of which appear to repeat, and only 92 have been properly localized, with redshift measurements \cite{xu2023blinkverse}. The largest publicly available sample is provided by the Canadian Hydrogen Intensity Mapping Experiment (CHIME) Collaboration \cite{amiri2022overview}.

One point that remains unclear is whether all FRBs repeat, as confirmation requires prolonged follow-up observations of the same sky region. While the occurrence rate of non-repeating FRBs was expected to align with that of cataclysmic events or compact-object births, recent studies report a significantly higher detection rate \cite{ravi2019prevalence, yamasaki2024true}. This apparent inconsistency persists in the most recent CHIME/FRB catalog \cite{amiri2021first}, which continues to classify the majority of sources as non-repeating. This suggests that many so-called non-repeaters may be repeaters with emissions below current sensitivity thresholds or recurrence intervals longer than typical monitoring durations. The definitive confirmation that an FRB is truly non-repeating will likely only come from the detection of a counterpart consistent with a cataclysmic event, such as a supernova or a gamma-ray burst \cite{pleunis2021fast}. 

Recently, several studies have explored the use of machine learning to identify hidden repeaters among apparently non-repeating FRBs. Unsupervised methods, such as Uniform Manifold Approximation and Projection (UMAP), have successfully clustered known repeaters and revealed new candidates directly from burst parameters \cite{chen2022uncloaking}. Furthermore, topological data analysis has revealed structured groupings suggestive of distinct source populations \cite{bhatporia2023topological}. Supervised classifiers have identified features such as brightness temperature and spectral width as strong discriminants \cite{luo2023machine}, while hybrid approaches using t-SNE and UMAP with burst  or standardized data have further improved separation power \cite{yang2023classifying}. Graph-based models, such as the Minimum Spanning Tree, have also been employed to isolate repeater-rich clusters \cite{garcia2024separating}. These approaches demonstrate the potential of data-driven techniques to uncover hidden structure within the FRB population and guide targeted follow-up strategies. They also underscore the importance of developing robust classification frameworks and effective feature selection methods to reliably distinguish between repeating and apparently non-repeating sources. 

Spectral properties, such as bandwidth, spectral index, and spectral running, which are sensitive to the underlying emission mechanism and propagation effects, including scattering and scintillation in the interstellar and intergalactic medium, have consistently emerged as important discriminants between repeating and apparently non-repeating FRBs. Several recent studies have incorporated these features into both supervised and unsupervised analyses, revealing that repeaters tend to exhibit narrower and more structured spectra, while non-repeaters often show broader, irregular profiles \cite{sun2025exploring, pleunis2021fast, chen2022uncloaking, qiang2025unsupervised}. However, despite these indications of heterogeneity, there is currently insufficient observational support to define firm subclasses within the repeater population. The available samples remain limited and affected by strong instrumental selection biases, which hinder the development of a robust taxonomy. Given these limitations, our analysis treats repeaters as a single class, avoiding premature sub-classification until future systematic observations can provide more definitive distinctions.

This paper builds upon recent advancements in classification of FRBs by developing a systematic and interpretable machine learning framework designed to distinguish repeating from non-repeating sources. Our approach integrates three hybrid pipelines that combine dimensionality reduction with clustering -- PCA + k-means, t-SNE + HDBSCAN, and t-SNE + Spectral Clustering -- each evaluated under two feature configurations: one based solely on primary CHIME/FRB catalog observables, and another incorporating additional physically derived quantities. All pipelines are optimized through a grid search using a custom F\textsubscript{2}-based scoring function to adjust the model parameters. This methodology complements previous approaches by integrating enriched feature representations with systematic clustering evaluation and optimization, contributing to robust classification performance.

In Section~\ref{sec:data}, we describe the data used in this study, including catalog selection, preprocessing steps,  identification of primary features and the construction of derived features. In Section~\ref{sec:method}, we describe our methodology in detail  and implementation of our grid search. 
Section~\ref{sec:results} presents results and discussions comparing clustering performance, visualization diagnostics, and analysis of feature importance using PCA loadings, mutual information, and permutation-based F\textsubscript{2} impact, as well as the identification of new repeating FRB candidates.  
Finally, conclusions are presented in Section \ref{sec:conclusions}.

\section{Data and Preprocessing} \label{sec:data}
\subsection{Data Acquisition and Preprocessing}
The primary data set used in this work is the CHIME/FRB Catalog 1 (2021) \cite{amiri2021first}. This catalog contains more than 500 Fast Radio Bursts, including both repeating and apparently non-repeating sources, observed between July 2018 and July 2019. Data include a variety of astrophysical observables such as flux densities, burst widths, and spectral properties derived from best-fit burst models {~\cite{fonseca2024modeling}}. To ensure the reliability of the analysis, we first filter the dataset by removing the following sources that do not have flux measurements: FRB20190307A, FRB20190307B, FRB20190329B,
FRB20190329C, FRB20190531A, and FRB20190531B. Additionally, we exclude duplicate entries of known repeating sources and remove rows with missing or misformed values. 

In this work, we deliberately use the CHIME/FRB Catalog 1 (2021)\footnote{This catalog contains 474 one-off events and 62 repeat bursts from 18 repeater sources.} to develop and validate our classification methodology. Repeaters identified only in the 2023 catalog \cite{andersen2023chime} were treated as non-repeaters during training,  by design. This approach was intended to replicate the conditions prior to the release of the 2023 catalog, allowing classification based solely on early data without knowledge of their future behavior. After generating repeater-like candidates using Catalog 1 (2021) alone, we compared our predictions against the CHIME 2023 catalog. 

\subsection{Primary Observables}

Our analysis begins with a set of nine primary features directly extracted from the CHIME catalog, summarized in Table~\ref{tab:primary_features}.

These features were selected based on physical motivation and their interpretability in terms of intrinsic emission properties. Fluence ($F_\nu$), peak flux ($S_\nu$), burst width ($\Delta t$), scattering timescale ($\Delta t_{ST}$), and spectral index parameters ($\gamma$, $r$) capture key aspects of FRB morphology, spectral structure, and propagation effects. In particular, the spectral index $\gamma$,  which quantifies the power-law slope of the emission spectrum, and  the spectral running $r$, which describes how this slope changes with frequency, are strongly correlated and jointly reflect the underlying radiation mechanism and medium inhomogeneities \cite{sun2025exploring}. 
We use \texttt{low\_freq} since it provides a physically interpretable estimate of the lower bound of the emission band, supported by relatively stable measurements within  CHIME's frequency range. Although bandpass limitations near 400--800\,MHz may affect both ends, the lower edge (\texttt{low\_freq}) is  better constrained due to CHIME's higher sensitivity and more uniform response in the 400--600\,MHz range, whereas the upper end (above $\sim$750\,MHz) suffers from more pronounced roll-off and greater calibration uncertainties. The reference (or 'pivot') frequency used by FitBurst to compute the \texttt{peak\_freq}, \texttt{low\_freq}, and \texttt{high\_freq} features is approximately 400.1953~MHz~\cite{Fonseca2024}. Furthermore, we initially considered including \texttt{peak\_freq} as a proxy for the emission centroid, but subsequent evaluation showed that it introduced redundancy with other features and slightly reduced the F\textsubscript{2} score across multiple clustering methods. 

Overall, the selected features emphasize physically meaningful observables directly tied to the FRB emission process, rather than instrument-specific or highly model-dependent quantities.

\begin{table}[H]
\centering
\caption{Primary features extracted from the CHIME/FRB catalog.}
\label{tab:primary_features}
\begin{tabularx}{\columnwidth}{lX}
\toprule
\textbf{Feature} & \textbf{Description} \\
\midrule
\texttt{snr\_fitb} & Signal-to-noise ratio of the burst based on the FitBurst model \cite{fonseca2024modeling}. \\
\texttt{dm\_exc\_ymw16} & Extragalactic DM component, computed as the excess above the YMW16 Galactic model~\cite{yao2017new}. \\
\texttt{flux} & Estimated flux density in Jy. \\
\texttt{fluence} & Integrated fluence in Jy~ms. \\
\texttt{width\_fitb} & Temporal width of the burst (ms) from the FitBurst best fit \cite{fonseca2024modeling}. \\
\texttt{scat\_time} & Scattering timescale estimated by the FitBurst model (ms) \cite{fonseca2024modeling}. \\
\texttt{sp\_idx} & Spectral index assuming a power-law spectrum \cite{fonseca2024modeling}. \\
\texttt{sp\_run} & Spectral running, i.e., curvature of the spectrum \cite{fonseca2024modeling}. \\
\texttt{low\_freq} & Lowest frequency of detection (MHz). 
\\
\bottomrule
\end{tabularx}
\end{table}

\subsection{Derived Quantities}

To complement the catalog-provided observables, we compute six physically motivated derived quantities using standard cosmological and radiative relations, following the methodology described in \cite{zhu2023machine}. These quantities serve as proxies for the intrinsic energetics, luminosity, and coherence scale of FRB emission.

\begin{itemize}
    \item \textbf{Redshift} ($z$): The observed dispersion measure, $\mathrm{DM}_{\mathrm{obs}}$, consists of contributions from both local and extragalactic (EG) environments. This is expressed as:
    \begin{equation}\label{dmobs}
        \mathrm{DM}_{\mathrm{obs}} = \mathrm{DM}_{\mathrm{local}} + \mathrm{DM}_{\mathrm{EG}}(z),
    \end{equation}
    \noindent where the local component is given in terms of contributions from the Milky Way interstellar medium (ISM) and the halo surrounding our galaxy:
    \begin{equation}
        \mathrm{DM}_{\mathrm{local}} = \mathrm{DM}_{\mathrm{ISM}} + \mathrm{DM}_{\mathrm{halo}},
    \end{equation}
    \noindent while the extragalactic component includes contributions from the intergalactic medium (IGM) and the host galaxy of the FRB:
    \begin{equation}
        \mathrm{DM}_{\mathrm{EG}} = \mathrm{DM}_{\mathrm{IGM}} + \frac{\mathrm{DM}_{\mathrm{host}}}{(1+z)}.
    \end{equation}
    The redshift is estimated from the excess dispersion measure using the Macquart relation \cite{deng2014cosmological}, under the assumption of a fully ionized intergalactic medium. The intergalactic DM component is modeled as:
    \begin{equation}
    \mathrm{DM}_{IGM}(z) = \frac{3 c H_0 \Omega_b f_{IGM}~\chi}{8 \pi G m_p }  \int_0^z \frac{(1+z')\,dz'}{E(z')},
    \end{equation}
    where $f_{\mathrm{IGM}}$ and $\chi$ represent the fraction and ionized fraction of baryons in the IGM. The cosmic baryon density, the mass of the proton, and the speed of light are denoted by $\Omega_b$, $m_p$, and $c$, respectively. $E(z)$ captures the cosmological dependence of this equation through the dimensionless Hubble parameter given by\footnote{We assume a flat $\Lambda$CDM cosmology throughout this work, specifically adopting the Planck 2018 best-fit parameters \cite{Planck2018}.} $E(z) = \sqrt{\Omega_m (1+z)^3+\Omega_\Lambda}$. We numerically invert this relation to estimate the redshift from \texttt{dm\_exc\_ymw16} for each burst. The values for the constants used in this work are listed in Table \ref{tab:constants}. The values adopted for $\mathrm{DM}_{\mathrm{halo}}$ and $\mathrm{DM}_{\mathrm{host}}$ were $30~\mathrm{pc}~\mathrm{cm}^{-3}$ and $70~\mathrm{pc}~\mathrm{cm}^{-3}$, respectively \cite{xu2015extragalactic,yamasaki2020thegalactic,dolag2015constraints,hashimoto2020noredshift,li2020cosmology}.

    \item \textbf{Frequency width} ($\Delta \nu$): The effective bandwidth is approximated as the difference between the highest and lowest  observed frequencies, scaled by redshift:
    \begin{equation}
        \Delta \nu = \left(\nu_{\text{high}} - \nu_{\text{low}}\right)\left(1+z\right),
    \end{equation}
    \noindent where $\nu_{\text{high}}$ and $\nu_{\text{low}}$ are provided in the CHIME catalog.

    \item \textbf{The time width in the reference frame} ($\Delta t_r$): The intrinsic time width is computed from the observed time width and corrected for cosmic expansion:
    \begin{equation}
        \Delta t_r = \frac{\Delta t}{(1+z)}.
    \end{equation}

    \item \textbf{Isotropic-equivalent energy} ($E$): The isotropic-equivalent energy emitted by the burst is estimated as:
    \begin{equation}
        E = 4 \pi D_L^2(z) \cdot F_\nu \cdot \Delta \nu \cdot (1+z)^{-1},
    \end{equation}
    where $F_\nu$ is the fluence in $\mathrm{Jy}\cdot\mathrm{ms}$, $\Delta \nu$ is the bandwidth in Hz and $D_L$ is the luminosity distance derived from the redshift.

    \item \textbf{Luminosity} ($L$): The burst luminosity is defined as the energy emitted per unit time:
    \begin{equation}
        L =  4 \pi D_L^2(z) S_{\nu} \nu_c,
    \end{equation}
    \noindent where $S_\nu$ is the peak flux and $\nu_c$ is the observed peak frequency.

    \item \textbf{Brightness temperature} ($T_B$): The effective brightness temperature is computed as \cite{luo2023machine}:
    \begin{equation}
        T_B = \frac{S_{\nu} D_L^2}{2 \pi k_B\left(\nu_c\Delta t\right)^2\left(1+z\right)},
    \end{equation}
    where $S_{\nu}$ is the peak flux density in Jy, $k_B$ is the Boltzmann constant, and the other parameters defined as above.
\end{itemize}


\begin{table}[H]
\centering
\caption{Physical constants and cosmological parameters adopted in this work.}
\label{tab:constants}
\begin{tabularx}{6.1cm}{l l X}
\toprule
\textbf{Constant} & \textbf{Value} & \textbf{Unit} \\
\midrule
c & $2.998 \times 10^8$ & m/s \\
G &  $6.674 \times 10^{-11}$ &  m$^3$/kg/s$^2$\\
$k_B$ & $1.381 \times 10^{-23}$ & J/K  \\
 $m_p$ & $1.673 \times 10^{-27}$ & kg \\ $H_0$ & $67.4$ & km/s/Mpc \\ $\Omega_m$ & 0.315 & -- \\
$\Omega_\Lambda$ & 0.685 & --  \\ $\Omega_b$ & 0.049 & -- \\$f_{\mathrm{IGM}}$ & 0.83 & -- \\
$\chi$ & 7/8 & --\\
\bottomrule
\end{tabularx}
\end{table}


We organize our analysis around two complementary scenarios for feature selection:

\begin{itemize}
    \item \textbf{Primary Only:} This configuration includes the nine features provided by the catalog discussed previously (signal-to-noise ratio, flux, fluence, width, scattering time, spectral index, spectral running, excess dispersion measure, and lowest detected frequency). This scenario isolates the predictive power of directly observed quantities, independent of any physical modeling assumptions.
    
    \item \textbf{Primary + Derived:} In this extended configuration, we supplement the nine primary features with six additional derived variables: redshift, frequency width, time width, isotropic-equivalent energy, luminosity, and brightness temperature. These quantities are computed using standard astrophysical relations and cosmological assumptions. 
    While model-dependent, these derived quantities encode key physical constraints related to emission energetics and propagation physics, serving as proxies for intrinsic burst properties.
\end{itemize}

This dual approach allows us to evaluate the influence of incorporating physically derived parameters on the performance of unsupervised clustering algorithms. In both cases, the selected features are standardized using z-score normalization before being fed into dimensionality reduction and clustering pipelines. This normalization was performed using the \texttt{StandardScaler} class from the \textit{scikit-learn} library~\cite{pedregosa2011scikit}, which applies the z-score formula to each feature. In other words, it subtracts the mean and divides by the standard deviation, resulting in features with zero mean and unit variance. This step is important to prevent features with larger scales from disproportionately influencing the results. Also, logarithmic transformations were applied to \texttt{bc\_width}, energy, luminosity, \texttt{flux}, \texttt{fluence}, and brightness temperature in both feature sets to account for their wide dynamic ranges and improve numerical stability during training.

The full modeling process--including dimensionality reduction, grid search for clustering optimization, and evaluation of clustering performance--is described in the following sections.

\section{Methodology}\label{sec:method}

Unsupervised learning provides a natural framework for exploring latent structures within FRB populations without imposing predefined labels. In this section, we detail the methodological pipeline used to uncover meaningful groupings and identify repeater candidates based on observational features.

\subsection{Dimensionality Reduction and Clustering Algorithms}

To identify the structure within the FRB parameter space, we employ unsupervised machine learning techniques that combine dimensionality reduction with clustering. This approach allows us to visualize complex relationships and reveal natural groupings in the data. Beyond its utility for visualization, dimensionality reduction also improves the efficiency and effectiveness of learning algorithms by alleviating issues associated with high-dimensional feature spaces, such as increased sparsity, redundancy, and susceptibility to overfitting. Before clustering, the high-dimensional feature space is projected onto a two-dimensional subspace to simplify the structure while preserving key relationships. In this work, we consider two-dimensionality reduction methods:

\begin{itemize}
    \item \textbf{Principal Component Analysis (PCA):} PCA is a linear transformation technique that projects the data onto orthogonal axes (principal components), capturing directions with the highest variance. It is computationally efficient and preserves global structure, but may struggle to capture non-linear patterns in the data \cite{mackiewicz1993principal}.

    \item \textbf{t-distributed Stochastic Neighbor Embedding (t-SNE):} t-SNE is a non-linear dimensionality reduction method, particularly effective for visualizing local structures in high-dimensional data. Preserve neighborhood relationships by modeling pairwise similarities and projecting them into a lower-dimensional space. t-SNE is sensitive to hyperparameters such as perplexity and exaggeration, which control the balance between local and global structure preservation \cite{hinton2002stochastic, van2008visualizing}.
\end{itemize}

Once the data are projected onto a two-dimensional plane, we apply clustering algorithms to group FRBs with similar features. The clustering algorithms used in this study are:

\begin{itemize}
    \item \textbf{k-means clustering:} A classic centroid-based algorithm, first introduced in \cite{lloyd1982least}, which partitions data into $k$ clusters by minimizing the variance within each group. It performs best when clusters are approximately spherical and have similar densities and variances.

    \item \textbf{HDBSCAN (Hierarchical Density-Based Spatial Clustering of Applications with Noise):} A density-based algorithm that identifies clusters of varying shapes and densities. It is robust to noise and does not require the number of clusters to be specified in advance. HDBSCAN labels points in low-density regions as noise, making it suitable for irregular astrophysical populations \cite{campello2013density}.

    \item \textbf{Spectral Clustering:} This graph-based algorithm constructs a similarity graph among data points and uses the eigenvectors of the Laplacian to group the data into clusters. 
    It is particularly effective in detecting non-convex structures and complex cluster boundaries that k-means may fail to capture \cite{ng2001spectral}.
\end{itemize}

We then evaluate three combinations of dimensionality reduction and clustering, selected to balance interpretability, flexibility, and computational efficiency:

\begin{itemize}
    \item \textbf{PCA + k-means:} A fast, fully linear baseline that projects features with PCA and applies k-means to the first two principal components.

    \item \textbf{t-SNE + HDBSCAN:} A flexible, noise-aware configuration that non-linearly projects the data and identifies arbitrarily shaped clusters while filtering out low-density outliers.

    \item \textbf{t-SNE + Spectral Clustering:} A hybrid model that uses the expressive t-SNE projection followed by a graph-based clustering algorithm to capture nuanced separations.
\end{itemize}

Each combination is independently optimized through grid search to identify the optimal configuration. The goal is to assess whether natural groupings -- particularly between repeating and non-repeating FRBs -- emerge under different algorithmic assumptions.

\subsection{Grid Search and Custom Scoring} \label{grid}

To determine the optimal configuration for each algorithmic pipeline, we perform a grid search over the relevant hyperparameter space. This process aims to identify parameter combinations that maximize clustering quality with respect to distinguishing repeating from non-repeating FRBs, while also avoiding overfitting and over-fragmentation.

Each combination of dimensionality reduction and clustering -- namely, PCA + k-means, t-SNE + HDBSCAN, and t-SNE + Spectral Clustering -- has its own set of hyperparameters. For each method, we evaluate all permutations of the following parameter values:

\begin{itemize}
    \item \textbf{t-SNE parameters:}
    \begin{itemize}
        \item \texttt{perplexity} $\in \{30, 50\}$ -- controls the effective number of neighbors.
        \item \texttt{early\_exaggeration} $\in \{8, 12\}$ -- influences the tightness of clusters in the early optimization stages.
    \end{itemize}

    \item \textbf{HDBSCAN parameters:}
    \begin{itemize}
        \item \texttt{min\_cluster\_size} $\in \{10, 15, 20\}$ -- sets the minimum number of points required to form a cluster.
        \item \texttt{min\_samples} $\in \{1, 3, 5\}$ -- defines the minimum density threshold for a point to be considered a core.
    \end{itemize}

    \item \textbf{Spectral Clustering parameters:}
    \begin{itemize}
        \item Number of clusters $k \in \{2, 3, 4\}$.
        \item Assignment method $\in \{\texttt{kmeans}, \texttt{discretize}\}$  -- determines how clusters are extracted from the eigenvectors.
    \end{itemize}

    \item \textbf{k-means parameters:}
    \begin{itemize}
        \item Number of clusters $k \in \{2, 3, 4\}$.
    \end{itemize}
\end{itemize}

For each combination of parameters, the model generates cluster assignments for the input data. We then evaluate cluster quality using several standard classification metrics, computed by assigning a binary repeater label to each cluster based on the proportion of known repeaters it contains. This framework enables the calculation of precision, recall and the F\textsubscript{2}-score for each configuration.

The \textit{precision} is defined as
\begin{equation}
\text{Precision} = \frac{TP}{TP + FP},
\end{equation}
where \(TP\) and \(FP\) represent the number of true and false positive predictions, respectively. It quantifies the proportion of bursts predicted as repeaters that are indeed known repeaters, penalizing models that incorrectly classify non-repeaters.

The \textit{recall} is defined as
\begin{equation}
\text{Recall} = \frac{TP}{TP + FN},
\end{equation}
where \(FN\) is the number of false negatives. Measures the proportion of true repeaters successfully identified by the model, favoring configurations that minimize missed detections.

To balance these two metrics, we compute the F\textsubscript{2}-score, a weighted harmonic mean that prioritizes recall over precision:
\begin{equation}
F_2 = \frac{5 \cdot \text{Precision} \cdot \text{Recall}}{4 \cdot \text{Precision} + \text{Recall}}.
\end{equation}
This formulation is particularly appropriate for our astrophysical application, where failing to identify a potential repeater is more detrimental than overpredicting one.

For each combination of parameters, the model generates cluster assignments for the input data. We then evaluate the quality of the clustering using a custom objective function that balances three aspects: classification performance, cluster interpretability, and robustness to noise. The scoring function is defined as:
\begin{equation}
\text{Score} = F_2 - \frac{\alpha (n_c - 2)^2}{10} - \beta \cdot \frac{n_{\text{noise}}}{N},
\end{equation}
where $F_2$ represents the F\textsubscript{2}-score computed from binary predictions distinguishing repeaters from non-repeaters. The term $n_c$ denotes the number of identified clusters, while $n_{\text{noise}}$ corresponds to the number of data points labeled as noise - applicable only in the case of HDBSCAN. $N$ is the total number of samples in the data set. The parameter $\alpha=1.0$ controls the penalty associated with deviations from a binary classification (with the ideal case being $ n_c=2$), and $\beta=0.3$ penalizes the proportion of points left unclassified due to noise.


Clusters in which more than $15\%$ of the members  \cite{zhu2023machine} were tagged as known repeaters in the CHIME catalog were labeled as "repeater-dominant", and all points within such clusters were assigned a tentative repeater label under that method. This procedure applies regardless of the total number of clusters identified. For HDBSCAN, points assigned to cluster $-1$ (i.e., noise) were excluded from the evaluation and voting, as they were not confidently assigned to any group.

This metric was designed to promote solutions that (i) maximize the correct identification of repeaters, (ii) avoid overfragmentation of the dataset into too many clusters, and (iii) reduce sensitivity to outliers or sparsely populated regions. The weights $\alpha$ and $\beta$ were empirically chosen to balance the relative importance of interpretability and robustness without overpowering the F\textsubscript{2} contribution. By incorporating these penalties directly into the evaluation, our grid search selects models that are both effective and physically interpretable in the astrophysical context of the FRB classification.

All computational analyzes conducted in this study were implemented using the Python programming language. The core machine learning components, including dimensionality reduction, clustering algorithms, and performance evaluation metrics, were primarily built upon the \texttt{scikit-learn} library \cite{scikit-learn}. In addition, the \texttt{hdbscan} package \cite{mcinnes2017hdbscan} was employed for density-based clustering.

\section{Results and Discussions}\label{sec:results}
\begin{table*}[htb]
\centering
\caption{Clustering performance metrics for each combination of dimensionality reduction and clustering method, grouped by feature set. The custom score includes a penalty for noise and excessive cluster count.}
\label{results_metrics}
\begin{tabularx}{15.5cm}{l l c c c c}
\toprule
\textbf{Feature Set} & \textbf{Method} & \textbf{Precision} & \textbf{Recall} & \textbf{F\textsubscript{2} Score} & \textbf{F\textsubscript{2} Custom Score} \\
\midrule
\multirow{3}{*}{Primary Only} 
& PCA + k-means       & 0.61 & 0.77 & 0.73 & 0.73\\
& t-SNE + HDBSCAN     & 0.39 & 0.84 & 0.68 & 0.65 \\
& t-SNE + Spectral Clustering    & 0.42 & 0.87 & 0.72 & 0.72 \\
\midrule
\multirow{3}{*}{Primary+Derived} 
& PCA + k-means       & 0.36 & 0.95 & 0.71 & 0.71\\
& t-SNE + HDBSCAN     & 0.67 & 0.71 & 0.70 & 0.69\\
& t-SNE + Spectral Clustering   & 0.43 & 0.95 & 0.76 & 0.76\\
\bottomrule
\end{tabularx}
\end{table*}

\subsection{Clustering Performance}

We evaluated three unsupervised clustering pipelines: PCA followed by k-means, t-SNE followed by Spectral Clustering, and t-SNE followed by HDBSCAN. Each method was applied to two feature sets: one with the nine primary catalog observables, and the other with all fifteen features, including derived quantities, as presented in Section~\ref{sec:data}. Hyperparameters were optimized through grid search (see Section~\ref{sec:method}).

Table~\ref{results_metrics} summarizes the clustering performance obtained for each combination of dimensionality reduction and clustering method, evaluated separately  under two configurations: primary features only and primary + derived features. The reported metrics include precision, recall, F\textsubscript{2} score and the F\textsubscript{2} custom score. Among all configurations, the best performance was achieved by the t-SNE + Spectral Clustering pipeline applied to the full feature set, with an F\textsubscript{2} score of 0.76. This result outperformed both PCA + k-means (F\textsubscript{2} = 0.71) and t-SNE + HDBSCAN (F\textsubscript{2} = 0.70) using the same features. For primary features only, the best F\textsubscript{2} score was obtained with PCA + k-means (0.73), closely followed by t-SNE + Spectral Clustering (0.72). These results suggest that while Spectral Clustering is effective overall, PCA combined with k-means slightly outperforms it under the restricted feature configuration. In other words, although non-linear dimensionality reduction methods can capture more complex structures, PCA still provides a robust performance when only primary features are available. 

For PCA + k-means and t-SNE + Spectral Clustering, the number of clusters $k$ was selected through grid search in the range $k \in \{2, 3, 4\}$. In practice, the best-performing configurations almost always resulted in $k = 2$, which aligns with the binary classification goal of separating repeaters from non-repeaters. On the other hand, for t-SNE + HDBSCAN, the number of clusters is determined automatically by the algorithm based on the underlying density structure. In most cases, HDBSCAN tends to yield many clusters; therefore, the custom F\textsubscript{2} function was needed for this algorithm, according to our approach. Additionally, some data points were labeled as noise (cluster = $-1$), particularly in low-density regions. These points were excluded from the evaluation metrics, as described in Section~\ref{grid}.

Furthermore, these results demonstrate the benefit of incorporating additional physically motivated features that improve separability in feature space, as well as the advantage of using non-linear dimensionality reduction. The improvement in F\textsubscript{2} score when including derived quantities in most pipelines indicates that physical properties such as redshift and luminosity, for example, carry a discriminative potential to distinguish repeater behavior. In particular, these features trace extragalactic distances and energetics, potentially associated with progenitor environments or emission mechanisms. Moreover, the high recall scores -- especially in spectral clustering scenarios -- are particularly relevant for identifying FRB repeaters, since missing a true repeater is more costly than allowing a few false positives. These results support the interpretation that repeaters and non-repeaters arise from distinct astrophysical subpopulations.



\begin{figure*}[p]
    \centering

    \subfigure[PCA + k-means]{
        \includegraphics[width=0.80\textwidth]{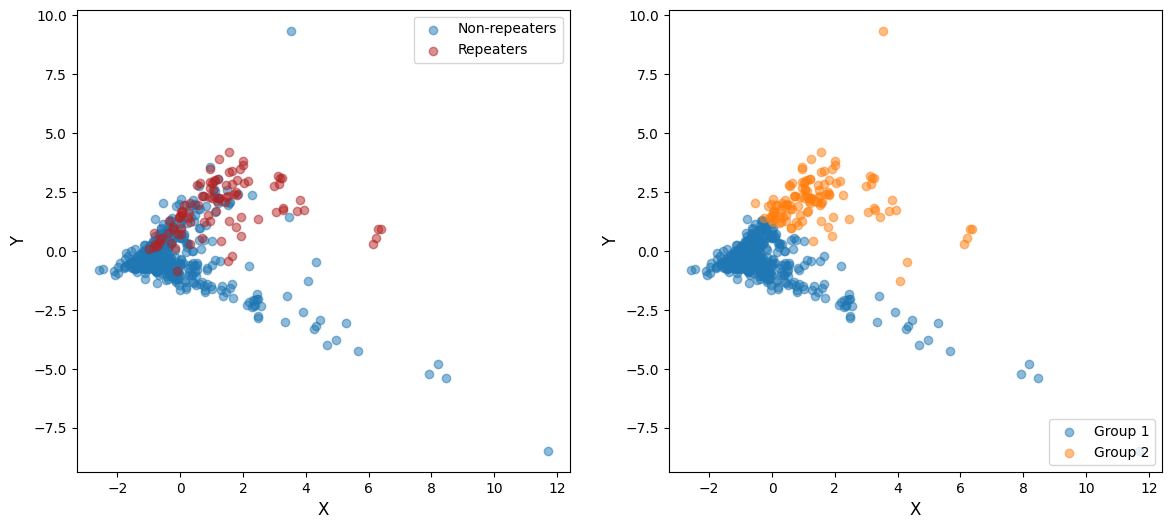}
        \vspace{0.5em}
    }

    \subfigure[t-SNE + Spectral Clustering]{
        \includegraphics[width=0.80\textwidth]{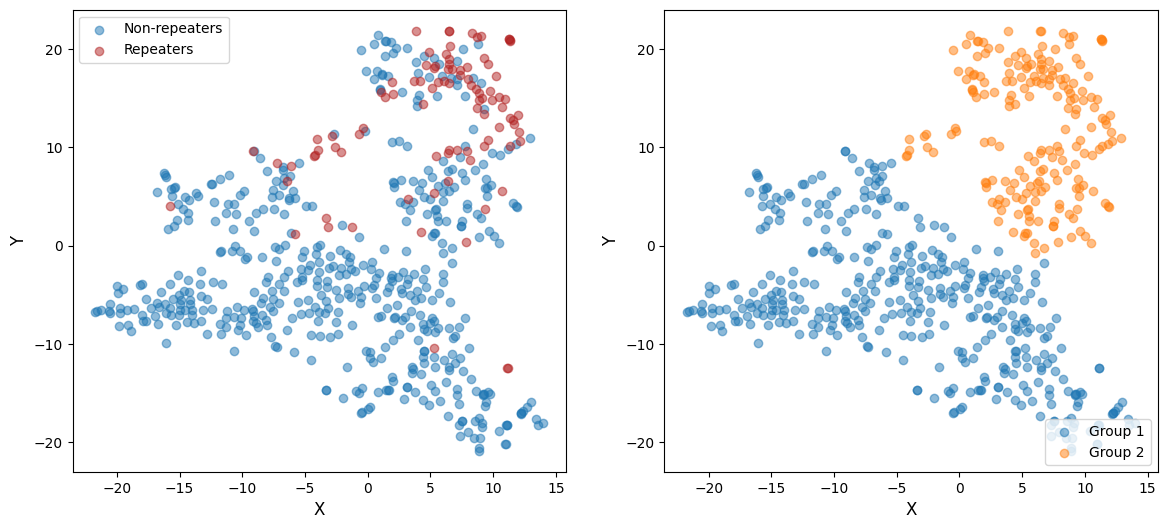}
        \vspace{0.5em}
    }

    \subfigure[t-SNE + HDBSCAN]{
        \includegraphics[width=0.80\textwidth]{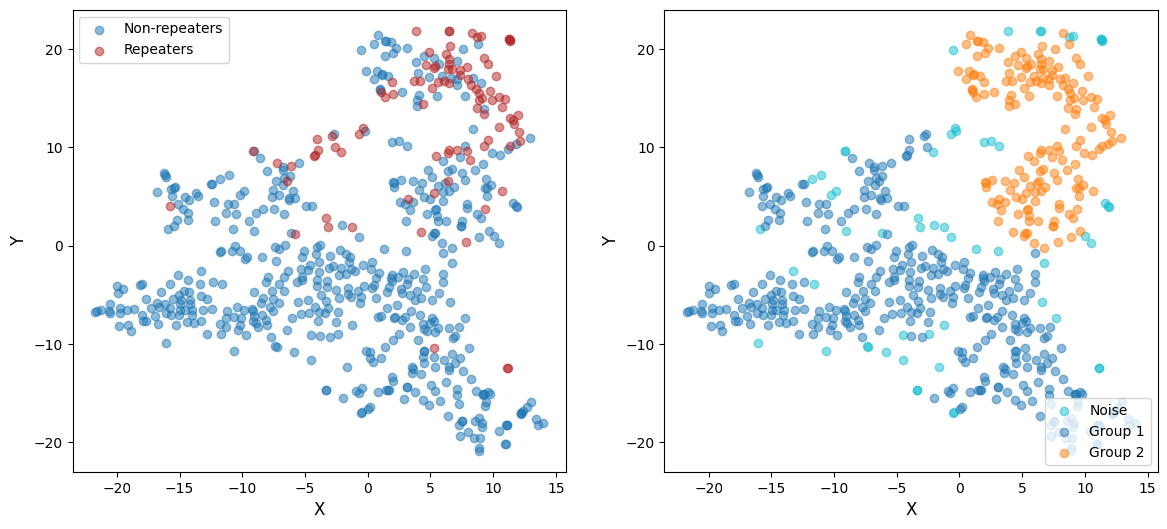}
    }

    \caption{Clustering visualizations using the \textbf{primary-only features}. Each panel shows a 2D projection of the FRBs colored by cluster assignment: (a) PCA + k-means, (b) t-SNE + Spectral Clustering, and (c) t-SNE + HDBSCAN.}
    \label{fig:clustering_primary}
\end{figure*}



\begin{figure*}[p]
    \centering

    \subfigure[PCA + k-means]{
        \includegraphics[width=0.80\textwidth]{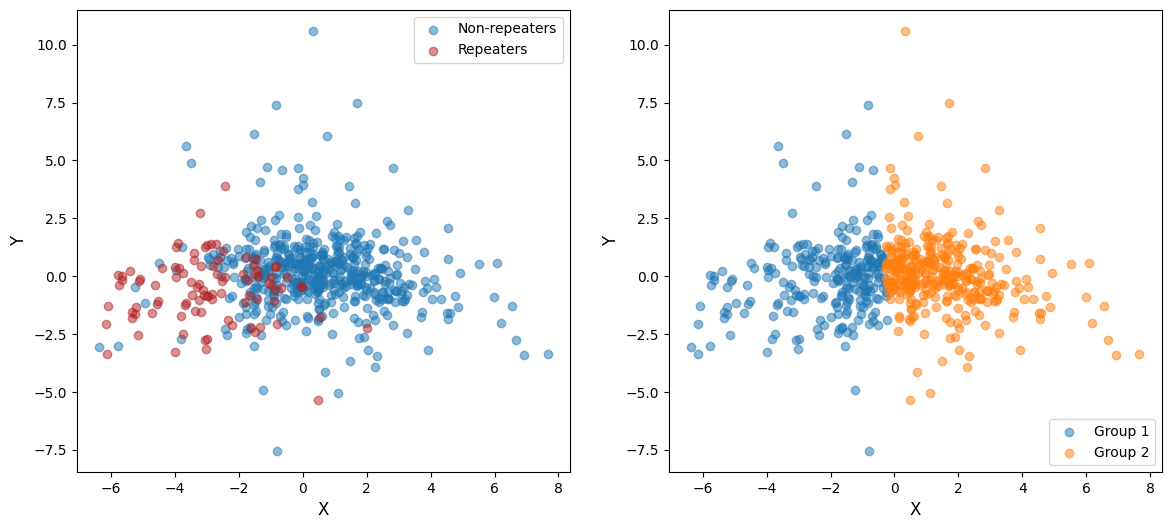}
        \vspace{0.5em}
    }

    \subfigure[t-SNE + Spectral Clustering]{
        \includegraphics[width=0.80\textwidth]{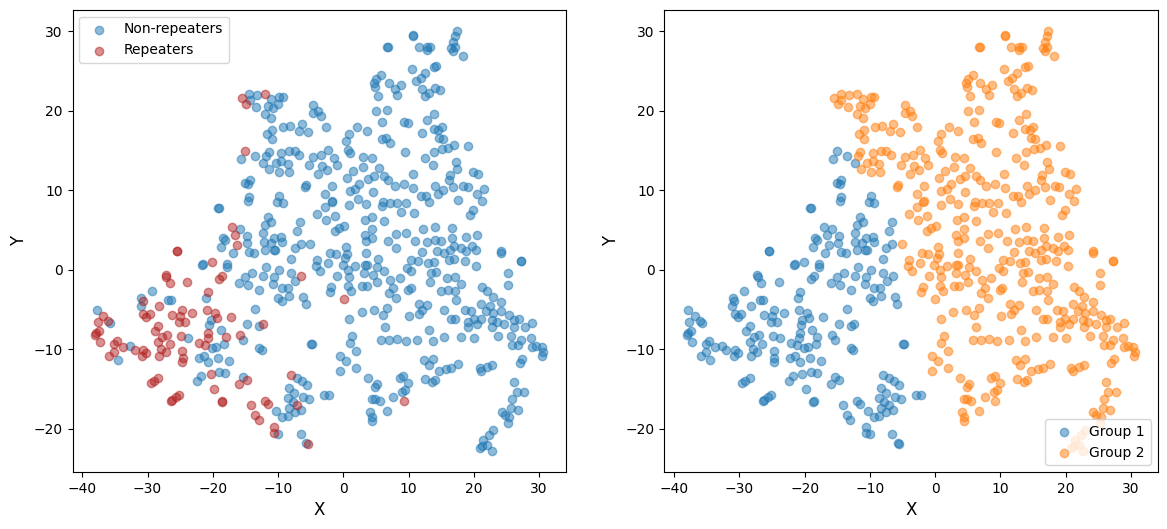}
        \vspace{0.5em}
    }

    \subfigure[t-SNE + HDBSCAN]{
        \includegraphics[width=0.80\textwidth]{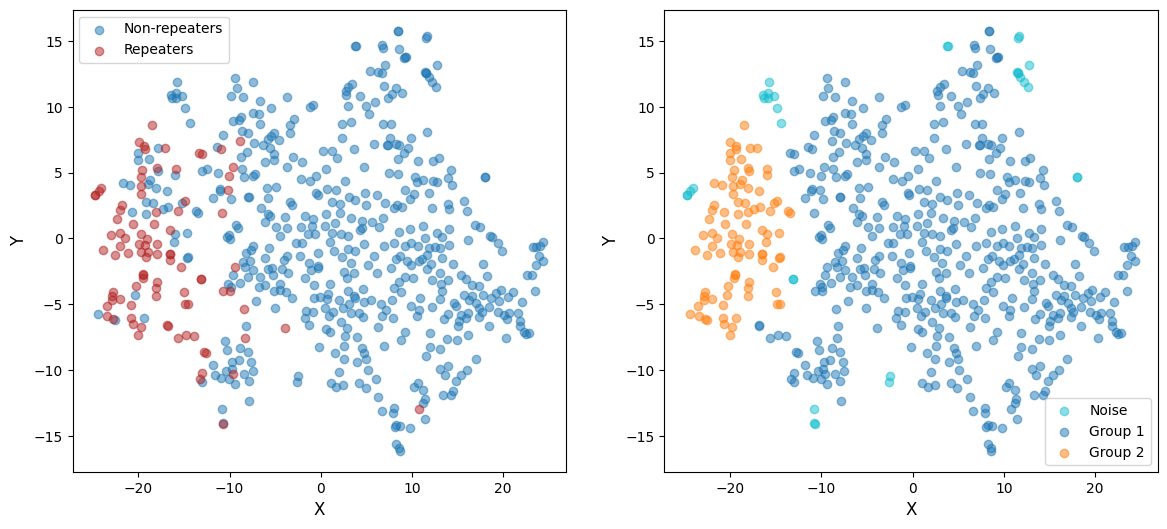}
    }

    \caption{Clustering visualizations using the \textbf{full feature set} (primary + derived). The panels show the same layout and clustering combinations as in Figure~\ref{fig:clustering_primary}, now using the extended set of physical quantities.}
    \label{fig:clustering_secondary}
\end{figure*}

In addition to quantitative metrics, we qualitatively examine the two-dimensional clustering results using scatter plots of the t-SNE and PCA projections. Figures~\ref{fig:clustering_primary} and~\ref{fig:clustering_secondary} present these visualizations for the primary and the primary + secondary feature sets, respectively. In the left panels of both figures, the true repeater labels -- flagged by CHIME -- are shown in red and blue, while the cluster assignments in the right panels are displayed in orange and blue.

For the primary-only case (Figure~\ref{fig:clustering_primary}), the t-SNE + Spectral Clustering combination (middle panel) exhibits the most coherent separation between repeaters and non-repeaters, with minimal overlap and compact group structure. PCA + k-means (upper panel) also achieves a degree of separation, although its linear projection limits cluster boundaries. The t-SNE + HDBSCAN model (lower panel) captures partial structure but is affected by noise points and more diffuse groupings. These results suggest that non-linear embeddings better capture the intrinsic structure of FRB properties that distinguish repeaters from one-off bursts, while density-based clustering appears more sensitive to noise, likely reflecting the variability inherent in FRB observations. 

When derived features are added (Figure~\ref{fig:clustering_secondary}), visual separability improves across all three models. The inclusion of physical quantities such as redshift and luminosity enhances manifold structure and highlights latent cluster shapes. Notably, the t-SNE + Spectral Clustering model (middle panel) maintains compact clusters and a clearer concentration of repeaters, consistent with its higher F\textsubscript{2} score. PCA + k-means (upper panel) also benefits from the extended feature set, producing nearly disjoint clusters. Meanwhile, t-SNE + HDBSCAN (lower panel) reveals a more coherent structure with fewer noise-labeled points. These improvements reinforce the astrophysical expectation that incorporating distance and energy indicators is essential for uncovering meaningful clustering of FRB sources.


\begin{figure}[p]
    \centering
    \subfigure[PCA + k-means]{
        \includegraphics[width=0.45\textwidth]{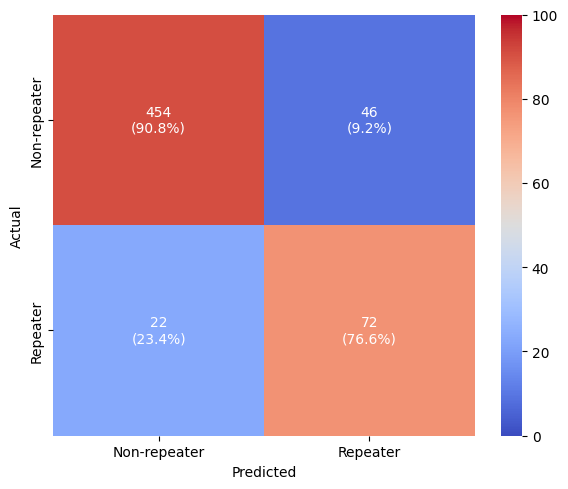}
    }
    \subfigure[t-SNE + Spectral Clustering]{
        \includegraphics[width=0.45\textwidth]{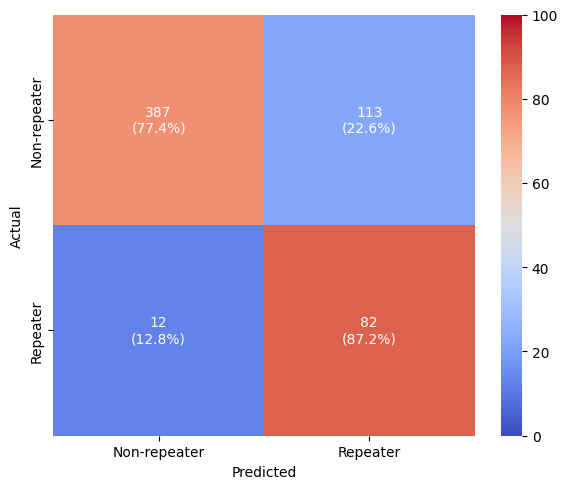}
    }
    \subfigure[t-SNE + HDBSCAN]{
        \includegraphics[width=0.45\textwidth]{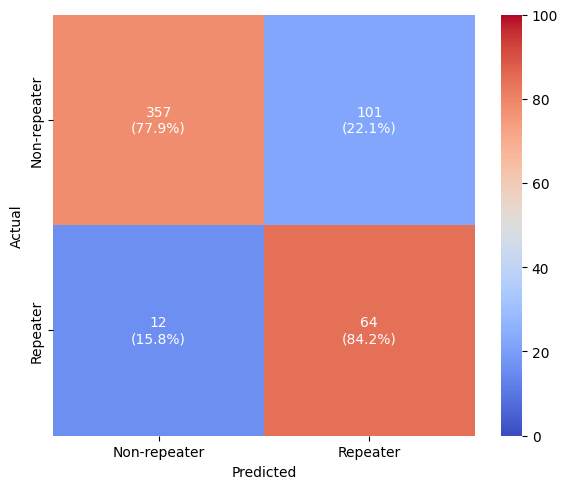}
    }
    \caption{Confusion matrices for the \textbf{primary-only feature set}. These plots show true vs. predicted repeater classifications for each clustering method. The t-SNE + Spectral Clustering configuration demonstrates clearer separation, while HDBSCAN provides good recall with moderate false positives.}
    \label{fig:conf_primary}
\end{figure}

\begin{figure}[p]
    \centering

    \subfigure[PCA + k-means]{
        \includegraphics[width=0.45\textwidth]{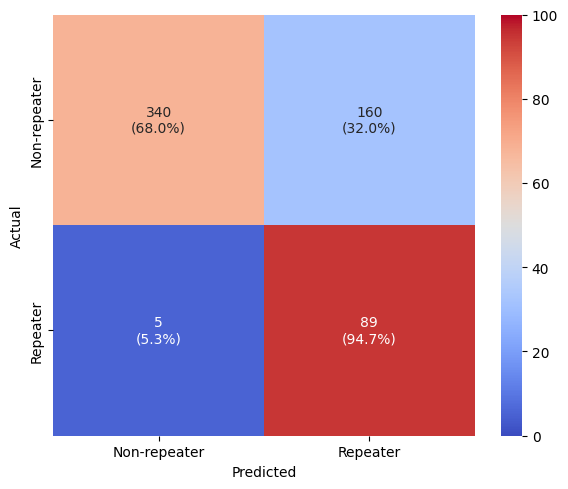}
    }
    \subfigure[t-SNE + Spectral Clustering]{
        \includegraphics[width=0.45\textwidth]{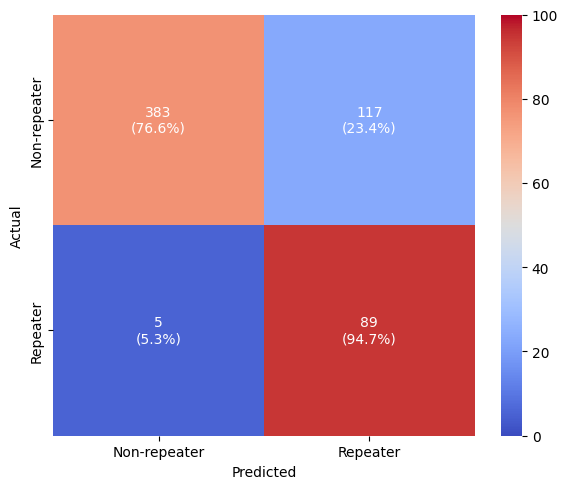}
    }
    \subfigure[t-SNE + HDBSCAN]{
        \includegraphics[width=0.45\textwidth]{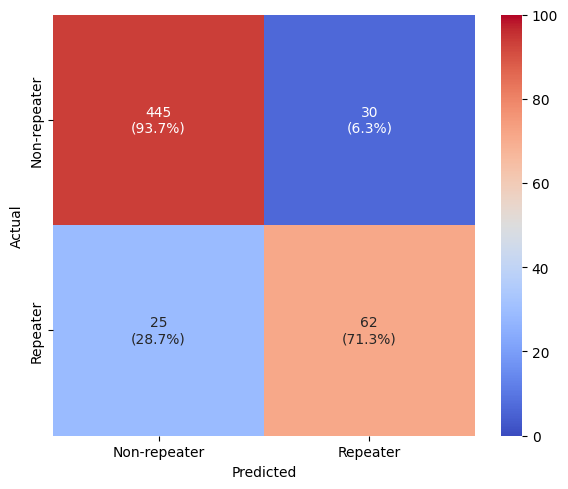}
    }
    \caption{Confusion matrices for the \textbf{full feature set} (primary + secondary). All three configurations show enhanced classification performance with derived quantities, particularly in the correct identification of repeaters.}
    \label{fig:conf_secondary}
\end{figure}

To visualize the classification performance of each clustering pipeline, we present confusion matrices for all three models using both the primary-only and full feature sets. Figures~\ref{fig:conf_primary} and~\ref{fig:conf_secondary} show the true vs. predicted repeater labels for optimal configurations selected via grid search.

For the primary-only case (Figure~\ref{fig:conf_primary}), the PCA + k-means configuration achieves a moderate balance between precision and recall, though there are some group confusion. The t-SNE + Spectral Clustering model improves both metrics, as seen in its denser diagonal entries. The t-SNE + HDBSCAN configuration yields slightly higher recall but introduces more noise-driven misclassifications. These results highlight the trade-off between precision and completeness across clustering approaches, a critical consideration for identifying repeaters, given their rarity and the inherent observational uncertainties. Only HDBSCAN produces noise-labeled points (cluster = $-1$), which account for a small fraction of the data. We tested assigning these points as either repeaters or non-repeaters and found that the impact on the F\textsubscript{2}  score was minimal, with no change in the ranking of methods.

When derived features are added (Figure~\ref{fig:conf_secondary}), we observe clear gains in true positive detection. Notably, both t-SNE + Spectral Clustering and PCA + k-means achieve strong diagonal dominance, especially in correctly identifying repeaters. t-SNE + HDBSCAN performs well, but at the cost of a slightly higher false-positive rate. These visual insights align closely with the metric-based results from Table~\ref{results_metrics}, further validating the clustering assignments. The best-performing hyperparameters for each clustering pipeline were selected via grid search and are listed in Table~\ref{tab:best_hyperparams}.

\begin{table}[ht]
\centering
\caption{Best hyperparameters for each method obtained from grid search.}
\label{tab:best_hyperparams}
\begin{tabularx}{\columnwidth}{lX}
\toprule
\textbf{Primary Only} & \\
\midrule
PCA + k-means & $k = 2$ \\
t-SNE + HDBSCAN & perplexity = 50, exaggeration = 8, \\
                & min\_cluster\_size = 15, min\_samples = 5 \\
t-SNE + Spectral & perplexity = 50, exaggeration = 8, \\
                 & $k = 2$, assign = \texttt{discretize} \\
\midrule
\textbf{Primary + Derived} & \\
\midrule
PCA + k-means & $k = 2$ \\
t-SNE + HDBSCAN & perplexity = 50, exaggeration = 8, \\
                & min\_cluster\_size = 15, min\_samples = 1 \\
t-SNE + Spectral & perplexity = 30, exaggeration = 8, \\
                 & $k = 2$, assign = \texttt{discretize}\\
\bottomrule
\end{tabularx}
\end{table}

To verify the robustness of our methodology, we repeated the full pipeline 100 times using different random seeds (0–99), including all dimensionality reduction and clustering steps. For the primary-only configuration, PCA + k-means and t-SNE + Spectral Clustering yielded the most consistent performance, with mean base F\textsubscript{2}  scores of $0.69 \pm 0.03$ and $0.72 \pm 0.01$, respectively. t-SNE + HDBSCAN produced a slightly lower mean F\textsubscript{2}  score of $0.68 \pm 0.04$, with greater variability due to its sensitivity to local density. For the primary + derived configuration, t-SNE + Spectral Clustering remained the best-performing and most stable method ($0.77 \pm 0.02$), followed closely by t-SNE + HDBSCAN ($0.75 \pm 0.03$). In contrast, PCA + k-means showed minimal variability ($0.710 \pm 0.001$). These results reveal that the overall ranking of methods remains consistent across runs and confirm the reproductibility of  the proposed pipeline.

Furthermore, to evaluate the contribution of propagation-related features, we tested an alternative configuration that include only quantities intrinsic to  FRB emission (fluence, flux, burst width, scattering time, spectral index, spectral running, and low frequency). The goal was to assess whether intrinsic properties alone could support a reliable classification. As shown in Table~\ref{results_intrinsic}~(\ref{appendixb}), this reduced set yielded competitive results, with F\textsubscript{2} scores reaching up to 0.73 for PCA + k-means and 0.72 for t-SNE + Spectral Clustering. However, when compared to the full feature set in Table~\ref{results_metrics}, we observe that the inclusion of derived and propagation-related features (e.g., redshift proxies, spectral bandwidths) improves both overall performance and method consistency, particularly for the best-performing t-SNE + Spectral pipeline.

\subsection{Feature Importance}

To identify which parameters contribute most to distinguishing repeaters from non-repeaters, we employed three complementary strategies: principal component analysis (PCA) loadings, mutual information (MI), and permutation importance based on the F\textsubscript{2} score. Each method offers a distinct perspective on feature relevance: PCA loadings reflect the contribution of each variable to the main axes of variance; mutual information quantifies the non-linear dependence between each feature and the repeater label; and permutation importance measures the drop in predictive performance when a feature is randomly shuffled.

\begin{figure}[ht]
  \centering
  \includegraphics[width=\linewidth]{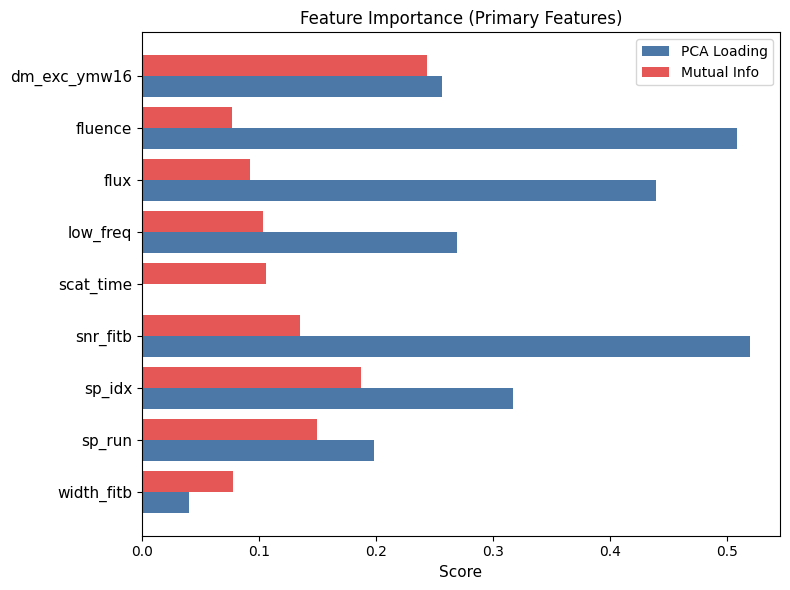}
  \caption{PCA loadings and mutual information scores for the \textbf{primary-only feature set}. The most influential features are related to signal intensity, propagation effects, and spectral shape.}
  \label{fig:pca_mi_primary}
\end{figure}

\begin{figure}[ht]
  \centering
  \includegraphics[width=\linewidth]{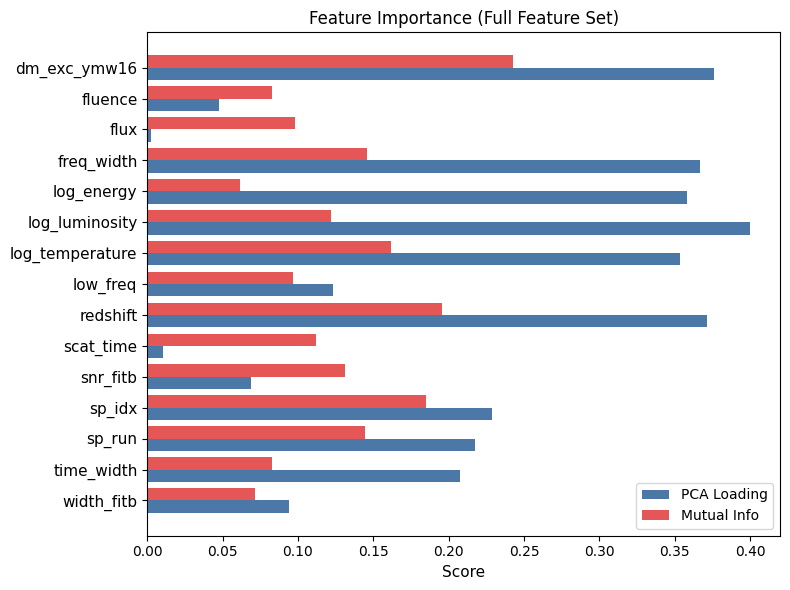}
  \caption{PCA loadings and mutual information scores for the \textbf{full feature set} (primary + derived). The top contributors are \texttt{log\_luminosity}, \texttt{dm\_exc\_ymw16}, and \texttt{redshift}.}
  \label{fig:pca_mi_full}
\end{figure}

Figures~\ref{fig:pca_mi_primary} and~\ref{fig:pca_mi_full} show the PCA loadings and mutual information scores for the primary-only and full feature sets, respectively. In the primary-only case, PCA loadings reveal that intrinsic properties like \texttt{snr\_fitb}, \texttt{fluence}, and \texttt{flux} dominate the first principal component, indicating that signal strength and burst amplitude are the main sources of variation. However, when derived features are included, the variance shifts toward properties such as \texttt{log\_luminosity}, \texttt{dm\_exc\_ymw16}, and \texttt{redshift}, emphasizing their importance in explaining the underlying FRB population structure. 

The results of mutual information align with these findings: \texttt{dm\_exc\_ymw16} and \texttt{sp\_idx} emerge as key predictors in both settings, while \texttt{redshift} and \texttt{log\_temperature} gain importance when derived features were included. The consistent presence of \texttt{sp\_idx} (spectral index) as a discriminative variable reinforces the idea that the spectral behavior of bursts plays a central role in repetition. These results support the interpretation that a combination of propagation and intrinsic properties influences repetition behavior.

\begin{figure}[ht]
  \centering  \includegraphics[width=\linewidth]{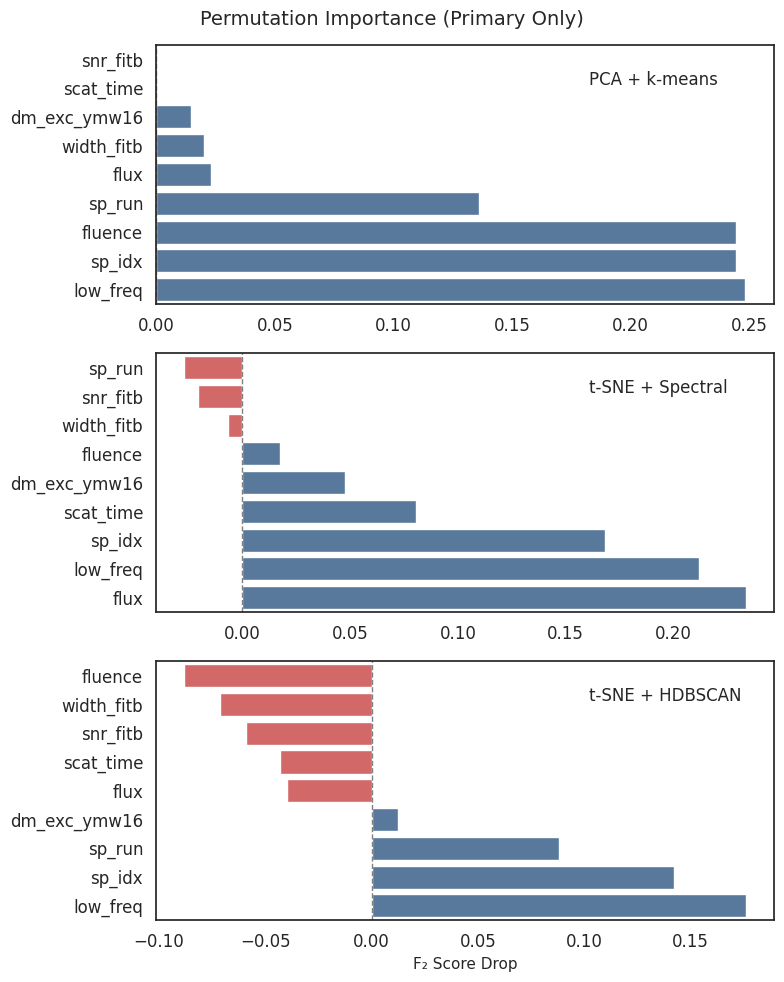}
  \caption{Permutation importance (F\textsubscript{2} drop) for the \textbf{primary-only feature set} using t-SNE + HDBSCAN, t-SNE + Spectral Clustering, and PCA + k-means.}
  \label{fig:f2drop_primary}
\end{figure}

\begin{figure}[ht]
  \centering
  \includegraphics[width=\linewidth]{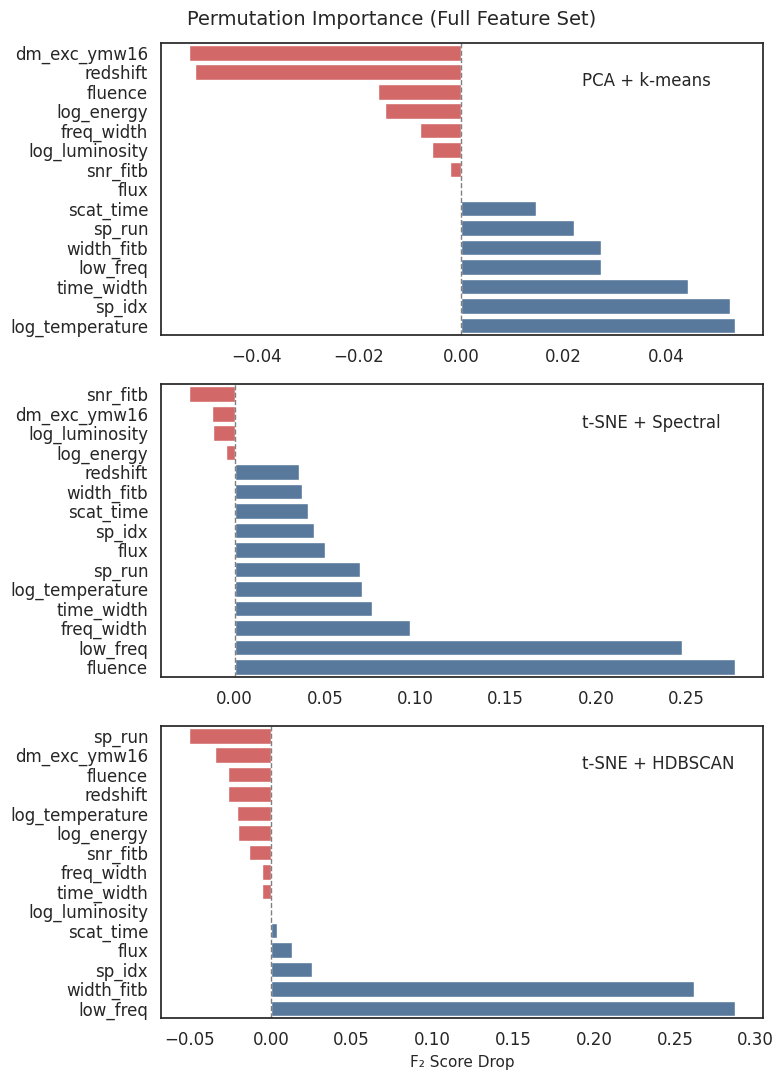}
  \caption{Permutation importance (F\textsubscript{2} drop) for the \textbf{full feature set} using t-SNE + HDBSCAN, t-SNE + Spectral Clustering, and PCA + k-means.}
  \label{fig:f2drop_full}
\end{figure}

To evaluate the impact of each feature on classification performance, we computed permutation importance by measuring the drop in the F\textsubscript{2} score when each feature was shuffled independently. This analysis was performed for each of the three clustering pipelines. Figures~\ref{fig:f2drop_primary} and~\ref{fig:f2drop_full} summarize the F\textsubscript{2} degradation for the primary-only and full feature sets, respectively. 

In the primary-only case, \texttt{low\_freq}, \texttt{snr\_fitb}, and \texttt{sp\_idx} produced the greatest F\textsubscript{2} drops, although the results were highly dependent on the clustering method. For instance, in the PCA + k-means pipeline, \texttt{fluence} and \texttt{low\_freq} have the strongest impact, while for t-SNE + HDBSCAN, \texttt{low\_freq} and \texttt{sp\_idx} were most impactful. With the full feature set, the most relevant variables varied across models: \texttt{width\_fitb}, \texttt{time\_width}, and \texttt{redshift} stood out in the HDBSCAN model, whereas \texttt{fluence}, \texttt{low\_freq}, and \texttt{dm\_exc\_ymw16} led in the spectral clustering configuration. These findings highlight the complex, model-dependent nature of feature relevance, but consistently reaffirm the central role of \texttt{dm\_exc\_ymw16}, \texttt{low\_freq}, and time-resolved or distance-based quantities. 

In some cases, reshuffling certain features led to a slight increase in the F\textsubscript{2} score, particularly for the t-SNE + HDBSCAN pipeline. However, when those same features were completely removed and the analysis was repeated, the overall performance dropped. This suggests that the features do contain relevant information, but may interact non-linearly with other features in the embedding space, especially in methods that are sensitive to local density variations. Such behavior is more likely a consequence of the clustering algorithm's sensitivity to perturbed feature distributions than actual evidence that the reshuffled features are uninformative.

Importantly, our feature importance results agree with previous studies such as \cite{sun2025exploring}, which used unsupervised machine learning techniques to assess the discriminative power among CHIME FRBs. In particular, the high ranking of \texttt{sp\_idx} (spectral index) and \texttt{sp\_run} (spectral running) in both mutual information and F\textsubscript{2}-based permutation tests reinforces earlier findings that frequency-dependent spectral behavior encodes key information about repetition. These spectral parameters likely reflect intrinsic emission processes and propagation effects that differ between repeaters and non-repeaters, suggesting that machine learning models are naturally sensitive to such signatures. Their persistent importance in our results could indicate that the pipelines capture genuine physical features that distinguish FRB types.

\subsection{Candidate Repeaters}


To identify potential new repeating FRB sources, we implemented a voting scheme that combines the outputs of three unsupervised clustering pipelines: PCA + k-means, t-SNE + Spectral Clustering, and t-SNE + HDBSCAN. For each method, FRBs were grouped according to their cluster assignments. 

An FRB was considered a candidate repeater only if the three clustering methods labeled it as such. This criterion provides a strong consistency filter and reduces false positives. We applied this analysis to the two configurations considered in this study: (i) using only the primary observational features from the CHIME catalog, and (ii) using the full feature set, which includes both the primary and derived physical quantities described in Section~\ref{sec:data}.

Tables~\ref{ap-1} and \ref{ap:2} list the FRBs predicted to be repeaters in each configuration, with confirmed repeaters from CHIME/FRB (2023) Catalog \cite{andersen2023chime} highlighted in bold. From the primary-only configuration, we identified 37 candidate repeaters. The full feature set configuration yielded 41 candidates. We compared these predictions with the updated CHIME/FRB (2023) Catalog \cite{andersen2023chime}, which lists 25 new repeating sources, including six reclassifications of FRBs previously considered as non-repeaters.

Upon cross-checking, we verified that two of the FRBs identified in our primary-only candidate set are now confirmed repeaters in CHIME/FRB (2023) Catalog: \textbf{FRB20190110C}, and \textbf{FRB20190430C}. For the full feature set, three matches were found: \textbf{FRB20190113A}, \textbf{FRB20190226B}, and \textbf{FRB20190430C}. Considering both configurations, we correctly predicted four of the six reclassified FRBs by the CHIME/FRB (2023) catalog, as they appeared in repeater-dominant clusters across all methods considered. This agreement supports the credibility of our approach, given that these predictions were made independently of the updated classification and suggests that additional sources flagged by our models may also be repeaters awaiting confirmation.

\section{Conclusions} \label{sec:conclusions}

We presented a comparative study of unsupervised machine learning techniques aimed at classifying FRBs and identifying new candidate repeaters. Our methodology combines dimensionality reduction techniques -- PCA and t-SNE -- with clustering algorithms including k-means, Spectral Clustering, and HDBSCAN. We evaluated their performance across two feature configurations: one using only primary observables from the CHIME/FRB Catalog 1, and another combining these with a set of astrophysically motivated derived quantities such as redshift, isotropic energy, and luminosity.

The performance of each method was assessed using standard metrics, complemented by a custom scoring criterion based on the F\textsubscript{2} score that penalizes over-fragmentation and excessive noise. The best clustering performance was achieved by t-SNE + Spectral Clustering with the full feature set, supporting the view that physically informed features enhance the separability between repeaters and non-repeaters. More broadly, our results demonstrate that t-SNE-based approaches are particularly effective at capturing the complex, non-linear structure underlying the FRB parameter space. 

To identify and understand the key factors behind successful classification, we evaluated feature importance using three complementary techniques: PCA loadings, mutual information with the repeater label, and permutation importance based on F\textsubscript{2} score degradation. Features such as \texttt{dm\_exc\_ymw16}, \texttt{redshift}, and \texttt{sp\_idx} consistently emerged as highly informative, reinforcing previous findings and independently confirming the central role of spectral properties in distinguishing repeaters from non-repeaters {~\cite{sun2025exploring}}.

In addition, we proposed new candidate repeaters through a voting scheme across all clustering pipelines. FRBs that consistently appeared in repeater-dominant clusters but had not been previously labeled as repeaters were flagged as potential repeaters. 
This yielded 37 candidates when using only primary features, and 41 when including derived quantities. Upon cross-referencing with the CHIME/FRB (2023) catalog, we found that some of our predictions aligned with sources recently reclassified as repeaters, suggesting that our methodology effectively captured underlying patterns. 
Specifically, for the case using only primary features, the p-value associated with the overlap between our predictions and confirmed repeaters is $p=0.0746$, while for the case including both primary and derived features, the corresponding p-value is $p= 0.0104$ (see \ref{appendixa} for details). These results indicate that the observed agreement with the CHIME/FRB (2023) catalog is unlikely to arise from random clustering outcomes, particularly when derived features are considered. This provides quantitative support for the efficacy of our unsupervised methodology in identifying latent repeater behavior based solely on observed properties.

We note that our methodology assumes a binary separation between repeater-like and non-repeater-like clusters. However, the possibility of multiple subclasses within each category, such as different populations of repeaters or diverse non-repeating progenitors, cannot be excluded. Preliminary tests with relaxed scoring constraints occasionally yielded additional subclusters enriched in repeaters, suggesting potential structure beyond the binary framework. Although our present approach prioritizes interpretability and robustness, investigating these finer subdivisions represents a promising direction for future work as larger and more diverse FRB samples become available.

Taken together, these findings underscore the potential of unsupervised learning, especially when guided by astrophysically motivated features, to uncover latent structure in FRB populations and to support repeater classification in a data-driven yet physically grounded manner. The framework developed here is readily adaptable to future FRB catalogs and extended feature configurations, and offers a foundation for more refined approaches, including probabilistic clustering, semi-supervised models, and time-resolved analyses of burst activity.


\section*{Acknowledgements}

JASF acknowledges support from the National Research Foundation of South Africa. WSHR acknowledges partial support from FAPES and CNPq. We thank the anonymous referee for the insightful comments and helpful suggestions.

\section*{Data availability}
The data used in this study are available from the CHIME/FRB public catalog at \url{https://www.chime-frb.ca/catalog}.

\bibliographystyle{unsrt} 
\bibliography{example}

\appendix

\vspace{1cm}
\section{Candidate Repeater Tables \textcolor{red}{and p-values}} \label{appendixa}
To assess the statistical significance of our repeater-like classifications, we examined the overlap between our predicted candidates and the confirmed repeaters in the CHIME/FRB 2023 catalog. Among the 468 sources initially labeled as non-repeaters, our method identified 37 repeater-like candidates using only primary features and 41 using the full feature set (primary+derived). Of these, two of 37 (primary) and three of the 41 (full set) were confirmed as repeaters (see Tables below).

Assuming the null hypothesis that confirmed repeaters are randomly distributed among the 468 sources, the probability of finding at least two confirmed repeaters among 37 candidates identified using only primary features follows the hypergeometric distribution. Specifically, the p-value corresponds to the probability of obtaining at least two "successes" (i.e., confirmed repeaters) in $n=6$ draws without replacement from a population of 468 sources, with 37 classified as repeater-like. This yields a p-value of
\begin{equation}
p=P(X \geq 2) = 0.0746,
\end{equation}
indicating that the overlap is moderately significant.

For the full feature set, the probability of  finding at least three confirmed repeaters among the 41 candidates corresponds to a p-value of
\begin{equation}
p=P(X \geq 3) = 0.0104\,.
\end{equation}
This smaller value indicates a higher statistical significance, suggesting that the observed overlap between our candidate list and the confirmed repeaters is unlikely to be due to random chance.
\begin{table}[p]
\centering
\caption{Predicted repeater candidates using \textbf{primary features only} 
FRBs already confirmed as repeaters in CHIME/FRB (2023) Catalog are indicated in \textbf{bold}.}
\begin{minipage}[t]{0.48\textwidth}
\centering
\begin{tabular}{cc}
\hline
\textbf{FRB Source} & \textbf{FRB Source} \\
\hline
FRB20180725A & FRB20190101B \\
FRB20180801A & \textbf{FRB20190110C} \\
FRB20180916C & FRB20190112A \\
FRB20181017B & FRB20190125A \\
FRB20181117C & FRB20190129A \\
FRB20181129B & FRB20190130B \\
FRB20181203B & FRB20190206A \\
FRB20181213B & FRB20190211A \\
FRB20181221A & FRB20190218B \\
FRB20181223B & FRB20190228A \\
FRB20181228B & FRB20190329A \\
FRB20181231B & FRB20190410A \\
FRB20190423B & FRB20190422A \\
FRB20190429B & FRB20190428A \\
FRB20190519J & \textbf{FRB20190430C} \\
FRB20190601C & FRB20190527A \\
FRB20190609A & FRB20190605D \\
FRB20190621C & FRB20190623B \\
FRB20190701C & \\
\hline
\end{tabular}
\label{ap-1}
\end{minipage}
\end{table}
\hfill
\begin{table}
\caption{Predicted repeater candidates using 
\textbf{primary + secondary features}. 
FRBs already confirmed as repeaters in CHIME/FRB (2023) Catalog are indicated in \textbf{bold}.}
\begin{minipage}[t]{0.48\textwidth}
\centering
\begin{tabular}{cc}
\hline
\textbf{FRB Source} & \textbf{FRB Source} \\
\hline
FRB20180907E & FRB20190109A \\
FRB20180909A & FRB20190112A \\
FRB20180920A & \textbf{FRB20190113A} \\
FRB20180925A & FRB20190124E \\
FRB20181017B & FRB20190125A \\
FRB20181129B & FRB20190128C \\
FRB20181203B & FRB20190129A \\
FRB20181218C & FRB20190206B \\
FRB20181221A & FRB20190206A \\
FRB20181231B & FRB20190218B \\
FRB20190103B & FRB20190221B \\
FRB20190105A & \textbf{FRB20190226B} \\
FRB20190106A & FRB20190228A \\
FRB20190323D & FRB20190329A \\
FRB20190409B & FRB20190410A \\
FRB20190411C & FRB20190412B \\
FRB20190414B & FRB20190422A \\
FRB20190423B & FRB20190429B \\
FRB20190430A & \textbf{FRB20190430C} \\
FRB20190609A & FRB20190617B \\
FRB20190625A & \\
\hline
\end{tabular}
\label{ap:2}
\end{minipage}
\end{table}





\clearpage
\section{Results for the burst intrinsic properties} \label{appendixb}

\begin{table}[htb]
\centering
\caption{Clustering performance metrics for the alternative feature configuration focused only on intrinsic burst properties.}
\label{results_intrinsic}
\begin{tabularx}{15.5cm}{l l c c c c}
\toprule
\textbf{Feature Set} & \textbf{Method} & \textbf{Precision} & \textbf{Recall} & \textbf{F\textsubscript{2} Score} & \textbf{F\textsubscript{2} Custom Score} \\
\midrule
\multirow{3}{*}{Primary Only} 
& PCA + k-means       & 0.60 & 0.77 & 0.72 & 0.72\\
& t-SNE + HDBSCAN     & 0.40 & 0.87 & 0.71 & 0.71 \\
& t-SNE + Spectral Clustering    & 0.41 & 0.87 & 0.71 & 0.71 \\
\midrule
\multirow{3}{*}{Primary+Derived}
& PCA + k-means       & 0.46 & 0.97 & 0.79 & 0.79\\
& t-SNE + HDBSCAN     & 0.60 & 0.74 & 0.71 & 0.68\\
& t-SNE + Spectral Clustering   & 0.40 & 0.92 & 0.72 & 0.72\\
\bottomrule
\end{tabularx}
\end{table}

\end{document}